\documentclass[sigconf]{acmart}

\AtBeginDocument{%
  \providecommand\BibTeX{{%
    \normalfont B\kern-0.5em{\scshape i\kern-0.25em b}\kern-0.8em\TeX}}}

\copyrightyear{2023}
\acmYear{2023}
\setcopyright{licensedothergov}
\acmConference[PE-WASUN '23]{Proceedings of
the Int'l ACM Symposium on Performance Evaluation of Wireless Ad Hoc,
Sensor, \& Ubiquitous Networks}{October 30-November 3 2023}{Montreal, QC, Canada}
\acmBooktitle{Proceedings of the Int'l ACM Symposium on Performance
Evaluation of Wireless Ad Hoc, Sensor, \& Ubiquitous Networks (PE-WASUN '23),
October 30-November 3 2023, Montreal, QC, Canada}
\acmPrice{15.00}
\acmDOI{10.1145/3616394.3618270}
\acmISBN{979-8-4007-0370-6/23/10}

\usepackage{paralist}
\usepackage{subcaption}

\usepackage{algorithm}%
\usepackage{algorithmicx}%
\usepackage{algpseudocode}%
\algnewcommand\algorithmicforeach{\textbf{for each}}
\algdef{S}[FOR]{ForEach}[1]{\algorithmicforeach\ #1\ \algorithmicdo}

\begin{document}

\title{Towards Edge-Based Data Lake Architecture for Intelligent Transportation System}

\author{Danilo Fernandes}
\email{dfc@laccan.ufal.br}
\affiliation{%
  \institution{Federal University of Alagoas}
  \streetaddress{Av. Lourival Melo Mota, S/N, Tabuleiro do Martins}
  \city{Maceio}
  \state{Alagoas}
  \country{Brazil}
  \postcode{57072-970}
}

\author{Douglas L. L. Moura}
\email{douglas.moura@dcc.ufmg.br}
\affiliation{%
  \institution{Federal University of Minas Gerais}
  \streetaddress{Av. Pres. Antônio Carlos, 6627, Pampulha}
  \city{Belo Horizonte}
  \state{Minas Gerais}
  \country{Brazil}
  \postcode{31270-901}
}

\author{Gean Santos}
\email{gean.santos@laccan.ufal.br}
\affiliation{%
  \institution{Federal University of Alagoas}
  \streetaddress{Av. Lourival Melo Mota, S/N, Tabuleiro do Martins}
  \city{Maceio}
  \state{Alagoas}
  \country{Brazil}
  \postcode{57072-970}
}

\author{Geymerson S. Ramos}
\email{geymerson.ramos@inria.fr}
\affiliation{%
  \institution{Univ Lyon, Inria, INSA Lyon, CITI}
  \streetaddress{56 Bd Niels Bohr, 69100 Villeurbanne}
  \city{Villeurbanne}
  \country{France}
  \postcode{69100}
}


\author{Fabiane Queiroz}
\email{fabiane.queiroz@laccan.ufal.br}
\affiliation{%
  \institution{Federal University of Alagoas}
  \streetaddress{Av. Lourival Melo Mota, S/N, Tabuleiro do Martins}
  \city{Maceio}
  \state{Alagoas}
  \country{Brazil}
  \postcode{57072-970}
}

\author{Andre L. L. Aquino}
\email{alla@laccan.ufal.br}
\affiliation{%
  \institution{Federal University of Alagoas}
  \streetaddress{Av. Lourival Melo Mota, S/N, Tabuleiro do Martins}
  \city{Maceio}
  \state{Alagoas}
  \country{Brazil}
  \postcode{57072-970}
}

\renewcommand{\shortauthors}{Danilo Fernandes et al.}

\begin{abstract}
  The rapid urbanization growth has underscored the need for innovative solutions to enhance transportation efficiency and safety. Intelligent Transportation Systems (ITS) have emerged as a promising solution in this context. However, analyzing and processing the massive and intricate data generated by ITS presents significant challenges for traditional data processing systems. This work proposes an Edge-based Data Lake Architecture to integrate and analyze the complex data from ITS efficiently. The architecture offers scalability, fault tolerance, and performance, improving decision-making and enhancing innovative services for a more intelligent transportation ecosystem. 
We demonstrate the effectiveness of the architecture through an analysis of three different use cases: (i) Vehicular Sensor Network, (ii) Mobile Network, and (iii) Driver Identification applications.
\end{abstract}



\maketitle

\section{Introduction}    

Urbanization is a global phenomenon characterized by the growth and development of large cities resulting from migrating people from rural to urban areas. It involves concentrating economic activities and infrastructure in urban centers~\cite{seto2010}. Rapid urbanization has modernized many people's lives and brought numerous challenges to transportation systems. These challenges include air pollution, noise, traffic congestion, and accidents~\cite{zheng2014}. 
In this context, \textit{Intelligent Transportation System} (ITS) emerges as a potential solution to address the impacts of urbanization and improve the efficiency and safety of transportation systems~\cite{its2010}. ITS integrates advanced communication and information technologies with transportation infrastructure to support various applications and services, such as real-time traffic management, road safety, smart parking, and autonomous vehicles~\cite{zhu2019}.

These applications transmit data through various communication technologies and obtain them from heterogeneous data sources, such as vehicles, road infrastructure, and citizens. 
The rapid growth of ITS systems made analyzing and processing the data challenging for traditional data processing systems presenting a new Big Data scenario~\cite{Ramos2023}. However, solutions are emerging to provide a flexible and scalable data store to handle the new ITS Big Data~\cite{sawadogo2021}. 
To provide this support, the Logical Data Lake (DL) concept~\cite{Ramos2023} enables efficient management of ITS Big Data, allowing advanced analytics, inference, and making data-driven decisions. 
Despite their proven benefits, existing DL architectures suffer from certain limitations when applied in the context of ITS:
(i) Vehicles typically move at high speeds, resulting in frequent disconnections, especially when combined with reduced communication range; 
(ii) ITS data is dispersed and heterogeneous, which poses a significant challenge for data integration; 
(iii) Cloud architectures can suffer from high communication and computing overheads due to the concentration of data and processing in a single central server. 

To address these challenges, we propose a novel edge-based data lake architecture that provides an abstraction layer and enables efficient data integration, cleansing, and inference for decision-making applications in ITS. Multi-access Edge Computing (MEC) infrastructure provides processing and storage resources at the network edge, such as at the cellular base station~\cite{taleb217}. In our approach, we design a distributed architecture across servers located at the network edge and in the cloud, with each layer utilizing the data differently.
The main contribution of this research is a novel edge-based data lake architecture that offers a scalable and efficient solution to handle the diverse data requirements of ITS, enabling enhanced decision-making, improved operational efficiency, and innovative services for a more intelligent and connected transportation ecosystem.
In our approach, data ingestion is performed logically and allows the direct and transparent utilization of distributed data, resulting in more efficient use of resources.
We analyze three use cases. 
The first use case is a Vehicular Sensor Network (VSN) application, which relies on vehicle data to determine a subset of vehicles to act as aggregation points and perform data offloading. The second use case is a mobile network application, which uses data from mobile devices to optimize the handover process. Finally, the third use case is the driver identification application, which integrates data from different automotive sensors to identify individual drivers.
The results show our proposed architecture's effectiveness in providing seamless data integration, low-latency processing, and flexible support for advanced data analysis in transportation systems. 

\section{Related Work} \label{sec:related}

Big Data Analytics has the potential to revolutionize transportation systems, bringing significant improvements to traffic management, reducing congestion, and increasing road safety. 
However, many issues and challenges still need to be addressed to ensure the effective use of these technologies, such as effective management, storage, and processing of large and complex data sets. Existing architectures explore using advanced big data management technologies, such as cloud computing, data warehousing, and data lakes, to address these issues and handle this data efficiently and cost-effectively.

Guerreiro et al.~\cite{guerreiro2016} propose an ETL (Extract, Transform, and Load) architecture for big data analytics in ITS applications, addressing an application scenario on dynamic toll charging for highways. The proposed architecture can efficiently process and model large volumes of raw traffic data using Apache  Spark and  SparkSQL for data processing and MongoDB for data storage. 
Similarly, Ramos et al.~\cite{Ramos2023} present a zone-based data lake architecture for smart cities and governments. Their approach enables the ingestion and integration of heterogeneous data sources from IoT systems, social media, data streams, information systems databases, and Data Warehouses. Additionally, they integrate all data via a flexible metadata system, which enables manual data labeling, automatic metadata extraction, and data queries through the metadata.

Zhu et al.~\cite{zhu2019} provided a comprehensive survey of Big Data Analytics in ITS and introduced a three-layer architecture composed of a data collection layer, a data analytics layer, and an application layer. They also presented several use cases of Big Data Analytics in ITS, such as road traffic flow prediction, road traffic accident analysis, public transportation service planning, personal travel route planning, and others.

Darwish and Bakar~\cite{darwish2018} proposed a fog-based architecture for ITS big data real-time processing called RITS-BDA. The fog nodes can form clusters vertically or horizontally, sharing computational and storage resources.
Singh et al.~\cite{singh2020} proposed BlockIoTIntelligence architecture of converging blockchain and AI for the Internet of Things (IoT).
They present cloud, fog, edge, and device layers hierarchically and use Blockchain technology to mitigate the security and privacy issue and provide decentralized big data analysis.

Dang et al. \cite{dang2021} propose a zone-based data lake architecture for IoT, Small, and Big Data. The authors consider an analysis-oriented metadata model for data lakes that includes the descriptive information of datasets and their attributes and all metadata related to the machine learning analyses performed on these datasets. The authors implemented a web application of data lake metadata management that allows users to find and use existing data, processes, and analyses by searching relevant metadata stored in a NoSQL data store within the data lake.
They also present two real-world use cases: dataset similarity detection and machine learning guidance.

\begin{table}[htb]
    \centering
    \footnotesize
    \renewcommand{\arraystretch}{1.4}
    \caption{Comparison with other approaches in the literature.}
    \begin{tabular}{|l|c|c|c|c|}
        \hline
         \textbf{Work} & \textbf{Edge} & \textbf{Data} & \textbf{Virtual data} & \textbf{ITS}\\
          & \textbf{computing} & \textbf{variety} & \textbf{integration} & \textbf{use cases}\\
         \hline
         \hline
         Guerreiro et al.~\cite{guerreiro2016} & & & & \checkmark \\\hline
         Ramos et al.~\cite{Ramos2023} & & \checkmark & \checkmark & \\\hline
         Zhu et al.~\cite{zhu2019} & & \checkmark & & \\\hline
         Darwish and Bakar~\cite{darwish2018} & \checkmark & \checkmark & & \\\hline
         Singh et al.~\cite{singh2020} & \checkmark & \checkmark & & \\\hline
         Dang et al.~\cite{dang2021} & & \checkmark & \checkmark & \\\hline
         Our work & \checkmark & \checkmark & \checkmark & \checkmark\\
         \hline
         
    \end{tabular}
    \label{tab:comparison}
\end{table}

Table \ref{tab:comparison} presents a comparison of the existing architectures for big data ITS.
Although previous works have presented promising solutions for ITS big data, they often rely on centralized architectures based on cloud infrastructure. Furthermore, there is a lack of a deep discussion about the infrastructure level design and the exploration of potential use cases within the transportation ecosystem. Our study aims to introduce a novel data lake architecture for ITS that utilizes the capabilities of Edge Computing to address the limitations of previous works. 
The proposed architecture offers an abstraction layer and performs data lake operations at the network's edge in a virtualized and distributed manner, enabling scalability, efficiency, and transparency.
    
\section{Edge-Based Data Lake Architecture} \label{sec:architecture}

The data lakes play a crucial role in the overall architecture corresponding to a data storage and intelligence layer.
Dealing with heterogeneity in Big Data landscapes and scaling in distributed environments are the main features that make it well suited for ITS platforms \cite{Kirimtat2020SmartCitySurvey}.
This layer collects, catalogs, cleans, and transforms data from IoT Devices.
These tasks compose a process known as data wrangling, which enriches the data and makes it suitable for analysis.  \cite{Koehler2021DataWrangling}.
Nevertheless, the ultimate feature is providing data analytics, the core of the application layer.

Data management is always a challenge when facing data heterogeneity. Data Lakes addresse this issue by storing all collected data in raw format and providing an evolving framework for data structuring and refinement \cite{Ramos2023}. Such a methodology is known as the Zone-Based Data Model \cite{Giebler2019DL}. 
Data collected from external sources are logically assigned to an ingested data zone \cite{sawadogo2021}. 
We transform the data enhancing the quality; thus, the outcome can move into the following zones \cite{sawadogo2021}.
Such transformations run until the data reaches the maximum level of maturity at which it is ready to be analyzed by decision-makers \cite{Ramos2023}.  

A data governance layer traverses all the zones to assure data security, privacy, quality, and monitoring \cite{Giebler2019DL}. 
Data governance requirements are different for each zone.
At the ingestion zone, data must satisfy conditions for security and consistency. 
The distillation zone should enforce data privacy by controlling access.
The processing zone reinforces data quality, and the monitoring zone provides reports on the health of specific components and the general infrastructure of the data lake.
Data governance should increase and stack as the data follows through the ingestion to the insights zone.
However, this is only one of several variants of zone architecture. 
Such architectures generally differ in the number and characteristics of the zones \cite{sawadogo2021}.

We usually deploy data lakes in the cloud due to the considerable computing resources available and the elasticity in their allocation. Their costs are often tied only to resource demand, skipping the infrastructure maintenance costs. However, in an environment with a considerable volume of sensors constantly providing data and demand for real-time analysis, the latency in uploading it to the cloud can be a bottleneck. 
Nevertheless, given its more constrained computing assets, this does not justify deploying the data lake at the edge. Thus, the cooperative adoption of both approaches can balance the trade-off between physical resources and latency to minimize costs. 

\begin{figure}[hbt]
\centering
\includegraphics[width = \linewidth]{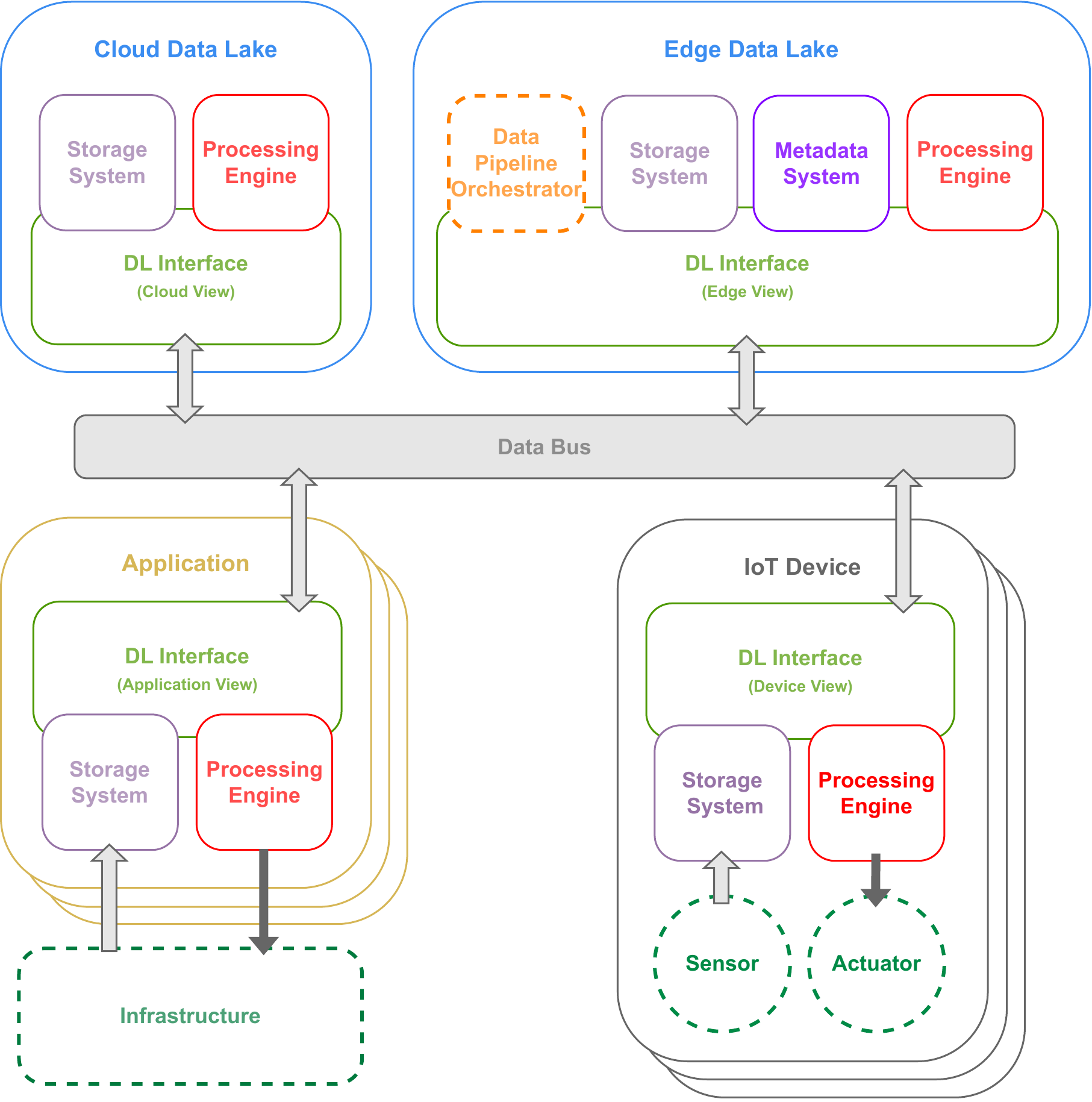}
\caption{Edge-based data lake architecture}
\label{fig:dl_architecture}
\end{figure}

In this light, the proposed data lake architecture is internally composed of two distributed data lakes running in the edge and cloud.
Although independent, these data lakes exchange data, operating together organically.
Our architecture (Figure \ref{fig:dl_architecture}) considers four layers: \texttt{IoT Device}, \texttt{Edge Data Lake}, \texttt{Cloud Data Lake}, \texttt{Application}. 
The \texttt{IoT Device} layer represents the mechanisms used to handle the data used by devices embedded in vehicles and other traffic-related devices. 
The \texttt{Edge Data Lake}, and \texttt{Cloud Data Lake} layers, collect, store, and sprocesses the data in batches or in real-time using distributed technologies at the edge or cloud. 
The \texttt{Application} layer uses the processed data. Examples of applications for ITS applications are congestion detection, traffic management, accident prediction, and other urban mobility solutions. Finally, all these layers exchange their data horizontally through a \texttt{Data Bus}. The \texttt{Data Bus} implements a lightweight messaging protocol to enable communication between all architecture components. An example of such a protocol is the \textit{Message Queuing Telemetry Transport} (MQTT)~\cite{pham2022}, which is designed to work on top of the TCP/IP protocol. This event-driven protocol acts on a publish/subscribe model, facilitating real-time communication within an ITS ecosystem.


\subsection{IoT Device Layer}\label{sec:sensors}

The \texttt{IoT Device} layer, also known as the perception layer, is where the smart sensors are in the data lake architecture. 
These devices may include various types of sensors, such as intra-vehicular sensors (e.g., chemical detectors, video cameras, and vibration/acoustic sensors), traffic control sensors (e.g., traffic lights, radar, inductive loops), and infrastructure sensors (e.g., roadside units and roadside weather stations). This layer collects data from the environment and makes it available to other layers for heavy processing and analysis. 
Wireless communication technologies like Bluetooth, Wi-Fi, Zigbee, LoRa, 5G, and others are commonly used in the device layer to transmit data.

In addition, the \texttt{IoT Device} layer has a \texttt{Processing Engine} to enable the performance of specific tasks and functions, such as data transformations, device actuator control, and communication. There is also a \texttt{Storage System} for local buffering of the data collected by the sensor and the processing outcome. This set of components can help reduce the amount of data that needs to be transmitted to the network and reduce bandwidth consumption. Finally, the DL interface handles all communication with the other layers.

\subsection{Edge and Cloud Data Lake Layers}\label{sec:dl}

The \texttt{Edge Data Lake} layer is mainly responsible for: \begin{inparaenum}[i)]
    \item Collecting data from sensors;
    \item Keep stored frequently used data (hot data);
    \item Data cataloging and metadata model inference;
    \item Real-time and reduced batch processing;
    \item Interfacing with MEC applications.
\end{inparaenum}
Additionally, the goal of the \texttt{Cloud Data Lake} layer is to extrapolate the computer power of the edge.
Then, we design it to: 
\begin{inparaenum}[i)]
    \item Storing huge volumes of sporadically used data (cold data);
    \item Costly batch processing.
\end{inparaenum}

The data kept by the ITS infrastructure should be in any layer present in the architecture.
The system sends the data from the lowest storage capacity tier to the highest as required for more costly processing or to free up storage space.
It executes the reverse direction when a layer with less computing power requests data stored in the next layer.
Consequently, hot data stays in the tiers closest to the application, while cold data accumulates in the cloud.

\texttt{Edge} and \texttt{Cloud Data Lakes} layers comprise \textit{Storage Systems}, \textit{Processing Engines}, and \textit{DL Interfaces}.
Nevertheless, the \texttt{Edge Data Lake layer} monitors and integrates the data, working as a virtual middleware in the data exchange of the other layers. 
For this purpose, the additional modules support the latter: \textit{Data Pipeline Orchestrator} and \textit{Metadata System}.
The \textit{Data Orchestrator Pipeline} is responsible for data ingestion from \texttt{Applications} and data exchange with \texttt{Cloud Data Lake}.
For the first case, the \textit{DL Interfaces} will intermediate the connection with the application, passing the received data (stream or batch) to the \textit{Orchestrator} and indicating a location in the \textit{Edge Storage System}.
When exchanging data with the \texttt{Cloud Data Lake}, the \textit{DL Interface} will inform the source and target data location in the cloud and edge \textit{Storage Systems}, and the \textit{Orchestrator} will transfer it in batches.
Note that such \textit{Orchestrator} is a distributed system; therefore, data transfer occurs in large volumes in parallel. 

The \textit{Storage System} stores the data on the edge. The \textit{Data Pipeline Orchestrator} and the \textit{Processing Engine} feed it.
In the last one, the stored data results from some processing task.
In addition, it returns data to all other components of the \texttt{Edge Data Lake} when requested via \textit{DL Interface}.
The \textit{Metadata System} is in charge of logically integrating all data and making it accessible to the user through homogeneous modeling. 
It also stores and manages information about the usage and performance of the data lake, for example, data lineage, history of data access, and data transfers.
When new data is ingested or created into the \texttt{Edge Data Lake}, it receives its location in the \textit{Storage System} and the available elementary metadata. 

In addition, it performs some processing and delegates the bulky ones to the \textit{Processing Engine} to extract relationships, hidden patterns, and semantic descriptors.
When the system transfers data from the edge to the cloud, we preserve the metadata, changing only its location to reflect its address in the cloud \textit{Storage System}.
In this way, this data remains ingested virtually at the \texttt{Edge Data Lake} and can be regressed when required.
All metadata is essential to empower data governance, enabling discovery, integrated queries, and unsupervised meta-analysis.
Note that big data implies big metadata, then \textit{Metadata System} pushes toward implementation in a distributed environment. Then, the metadata structure is across the computing nodes, which share the content with others when requested.

\begin{figure}[hbt]
    \centering
    \includegraphics[width = \linewidth]{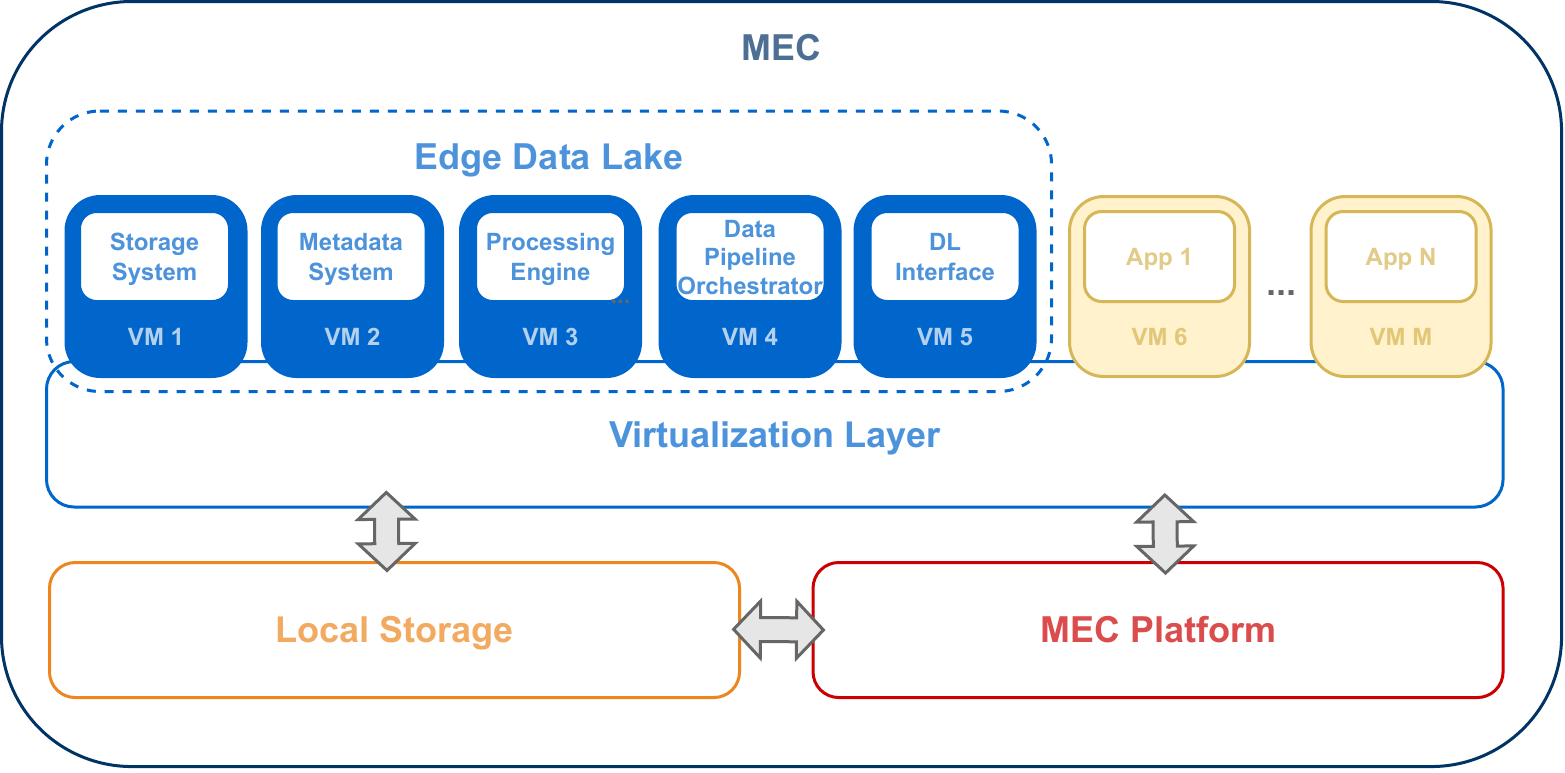}
    \caption{MEC Implementation}
    \label{fig:mec}
\end{figure}

The \textit{Processing Engine} receives requests from \textit{Metadata System} and \textit{DL Interface} to extract meta-features and refine data when the data evolves, the output returns to the requester or saves into \textit{Storage System}.
The \textit{DL Interface} communicates with the MEC \texttt{Applications} and translates the requests into coordinated tasks to the other modules within \texttt{Edge Data Lake}. 
\texttt{Edge Data Lake} runs distributed along the MEC servers, sharing computing resources with the applications. Figure \ref{fig:mec} presents a possible implementation of edge services on a MEC server.

All \texttt{Edge Data Lake} module nodes run side-by-side with MEC \texttt{Applications} in virtual machines (VM).
The \textit{Virtualization Layer} manages the virtual machines, which allocates resources to each of them and intermediates the communication between them and external ones within other MEC servers or clouds. 
The \textit{MEC Platform} consists of essential functionality required to run MEC applications. 
It can also enable MEC applications to provide and consume MEC services (e.g., radio network information, location, and traffic management services).
Finally, the Local Storage saves the sensing data, the VMs with their data, and the MEC operational data.

\subsection{Application Layer}\label{sec:apps}

The \texttt{Application} layer is responsible for providing valuable services to end-users based on the data processed by the data lake. Applications typically run on the MEC host but can also run in the cloud, on the device, or a combination. The choice of execution location depends on factors like the application's requirements, resource availability, and network conditions. These applications include typical ITS use cases, such as autonomous vehicles, smart parking, environmental monitoring, and video streaming. Other ones can also be run on this layer, such as data lake operations (e.g., distributed processing) and complex tasks (e.g., video processing, machine learning, augmented reality, and gaming) offloaded by IoT devices.

The \texttt{Application} has a \textit{Storage System} and \textit{Processing Engine} to perform their operations autonomously and a \textit{DL Interface} to communicate with the other layers of the architecture.
In addition, they can interact with the \textit{Infrastructure} that supports it by collecting and storing data in the \textit{storage system} and changing its setup via the \textit{Processing Engine}.
This capability enables, for example, the adaptation of network settings to optimally meet the demand for services.

\section{Use Cases of ITS Data Lake}\label{sec:cases}

\begin{figure*}[htb]
    \centering
    \begin{subfigure}{0.3\textwidth}
        \centering
        \includegraphics[width=\linewidth]{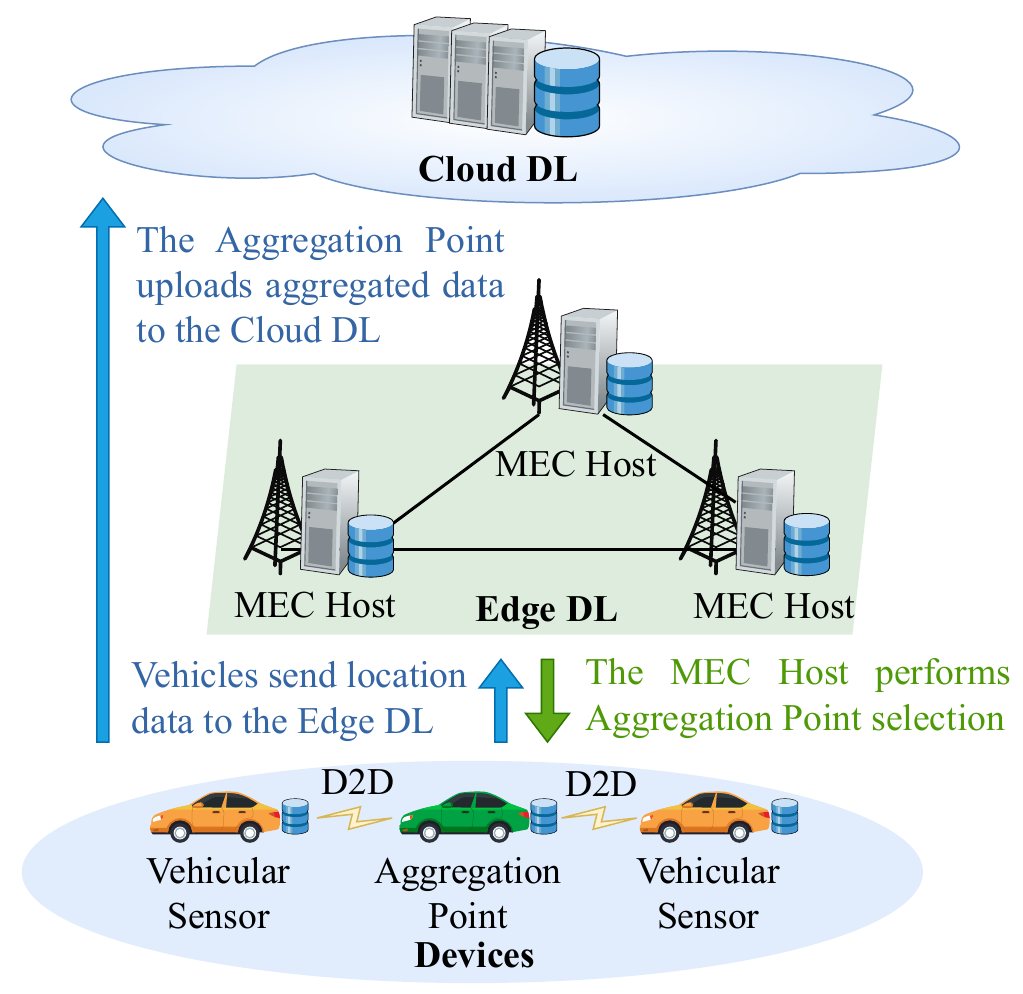}
        \caption{VSN application.}
        \label{fig:conceitual-centrality}
    \end{subfigure}
    \begin{subfigure}{0.3\textwidth}
        \centering
        \includegraphics[width=\linewidth]{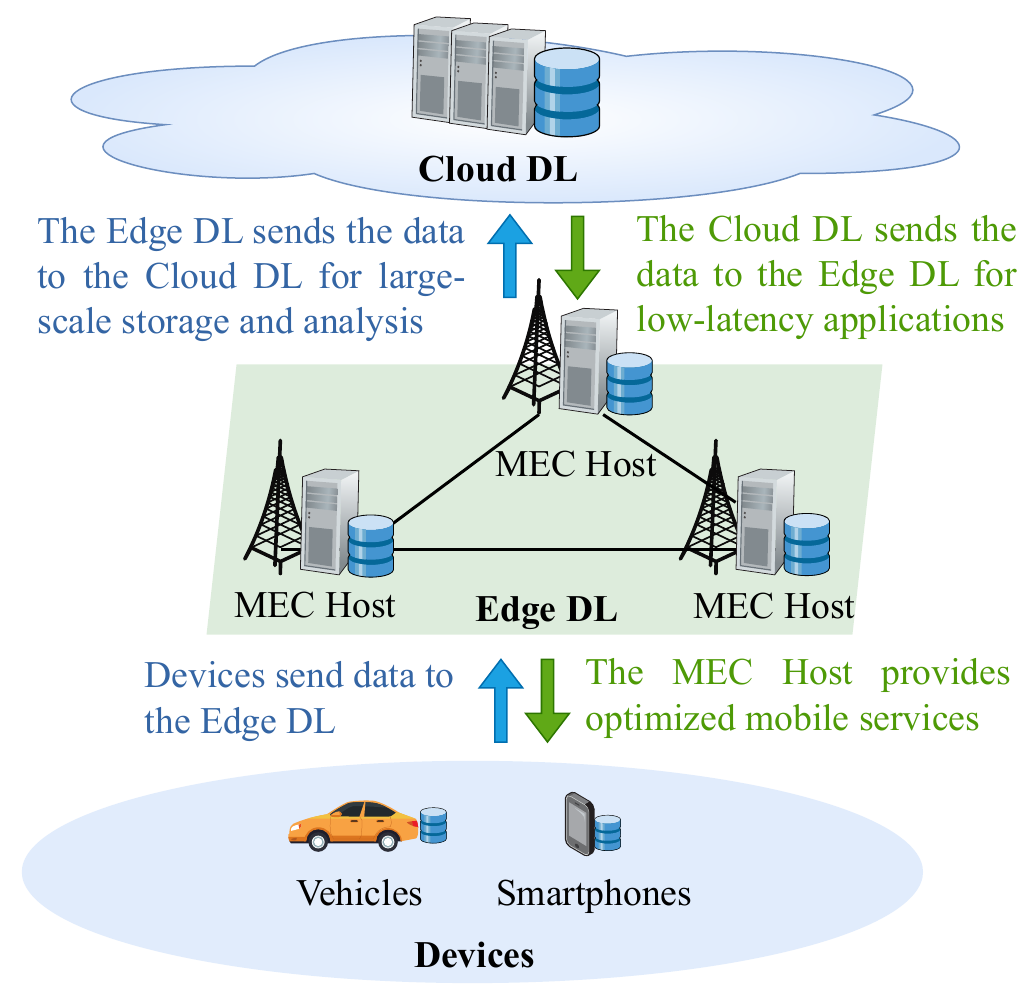}
        \caption{Mobile Network application.}
        \label{fig:mobile_app_framework}
    \end{subfigure}
    \begin{subfigure}{0.3\textwidth}
        \centering
        \includegraphics[width=\linewidth]{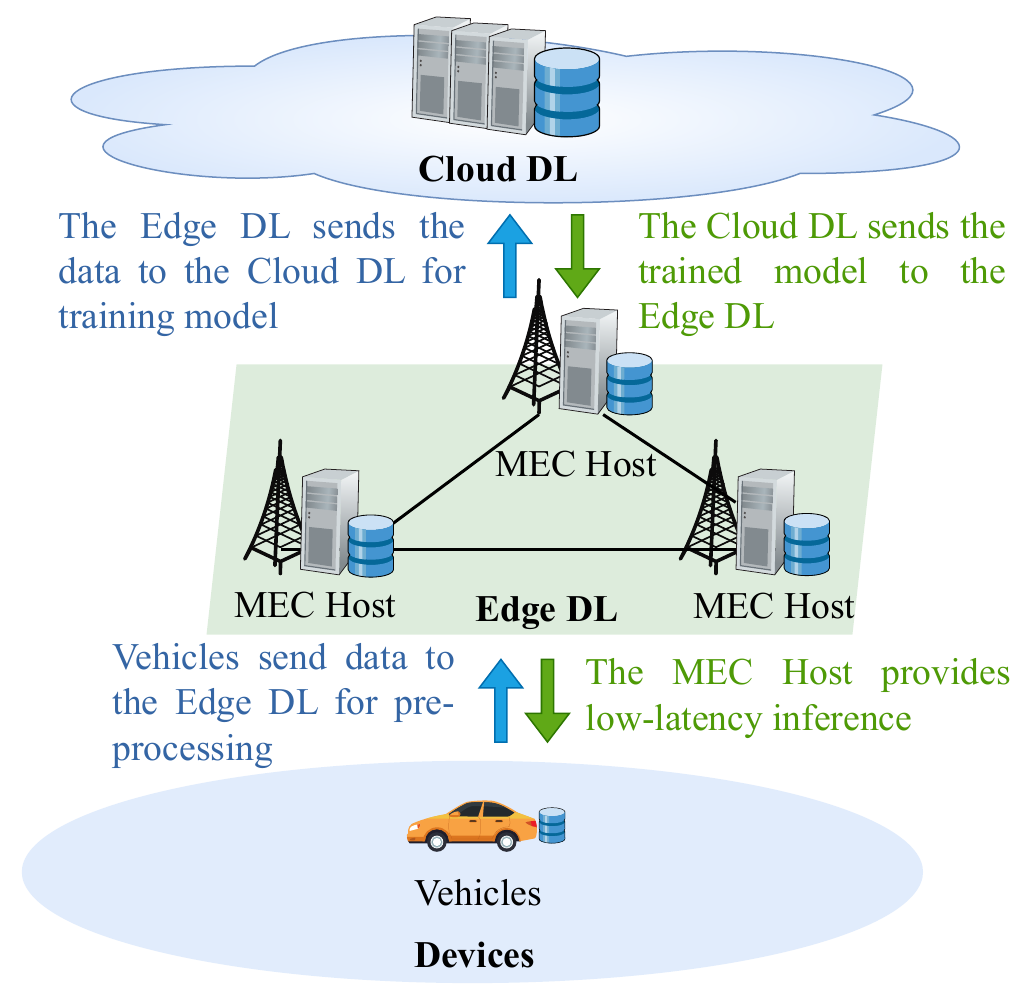}
        \caption{Driver identification application.}
        \label{fig:drive-identification-framework}
    \end{subfigure}
    \caption{Evaluated use case scenarios.}
    \label{fig:figura_completa}
\end{figure*}

In this section, we illustrate the potential of the proposed architecture by discussing three examples of use cases that demonstrate its benefits in ITS applications. 
We consider Vehicular, Mobile Network, and Driver Identification applications, illustrated in Figure~\ref{fig:figura_completa}.

\subsection{Vehicular Sensor Network Application}

Vehicular Sensor Network (VSN) refers to a promising paradigm of Remote Sensing, in which vehicles equipped with various sensing devices, powerful processing units, and wireless communication can act as mobile sensors and perform monitoring of the urban environment. 
VSN will enable various new ITS applications through the remote processing of data periodically collected by vehicles, such as improving road safety, traffic management, intelligent navigation, pollution monitoring, urban surveillance, and forensic investigations.

Many applications require a periodic upload of the sensory data transmitted to a monitoring center on the internet.
However, transmitting massive data over the cellular network can compromise network resources.
Data offloading is a potential solution to save bandwidth and prevent cell network overload, in which only a minimal number of vehicles, called aggregation points, will transmit data on the cellular network. An aggregation point is responsible for collecting sensory data generated by neighboring vehicles using Device-to-Device (D2D) communication, aggregating and uploading them. Our challenge is to determine which vehicles will act as aggregation points.

In this scenario, following the flow of Figure~\ref{fig:conceitual-centrality}, the vehicles represent the \texttt{IoT Devices} and generate sensing data with a high space-time correlation. We summarize these data through aggregation operators to combine data from several sources into a single value. First, vehicles share location information to the \texttt{Edge Data Lake} through the \textit{DL Interface}. The MEC Host represents the \texttt{Edge Data Lake}. Next, the \textit{Processing Engine} processes the raw data in parallel and transforms it into a suitable data format for storage in the \textit{Storage System}. The \textit{Metadata System} adds additional metadata information, such as creation date, data format, geographic location, and others. 
The MEC Host host the neighbor discovery service to provide a mechanism for devices to discover nearby devices.
A MEC application will run the \textbf{Centrality-based algorithm}~\cite{Moura2019} to solve the aggregation points selection problem. The algorithm uses the neighbor discovery service to model the graph and then selects the aggregation points. Finally, aggregation points collect sensory data from neighboring vehicles
using D2D communication. The aggregation point stores the data locally in the \textit{Storage System}, and the \textit{Processing Engine} performs data aggregation. The aggregated data is transmitted to the \texttt{Cloud Data Lake} for later analysis through the \textit{DL Interface}.

The proposed scheme has the following steps:
\begin{inparaenum}[i)]
    \item \textbf{Awareness} - The neighbor discovery service establishes the proximity relationship between the vehicles using the processed location data stored in the \texttt{Edge Data Lake};
    \item \textbf{Modeling} - The MEC \texttt{Application} model the  graph from the neighbor discovery service information;
    \item \textbf{Selection} - The Centrality-based algorithm uses closeness centrality in a greedy approach to select which vehicles will access the cellular uplink;
    \item \textbf{Upload} - Finally, a vehicle selected to access the cellular uplink receives a message from the MEC \texttt{Application} requesting the upload. The vehicle can then transmit the collected data.
\end{inparaenum}
The \textit{Data Pipeline Orchestrator} intermediates the data exchange with \texttt{Cloud Data Lake}. The sensory data is ingested into the \texttt{Cloud Data Lake} via an API, processed, and stored in the distributed file system with added metadata for organization and additional information.

We evaluate the proposed solution in a realistic simulation scenario derived from data traffic (TAPASCologne~\cite{Uppoor2014}) containing more than 700,000 individual car trips for 24 hours.
We compare our approach with a reservation-based algorithm~\cite{stanica2013}, and the results indicate an 
the aggregation rate improved by up to 10.45\% 

\begin{figure}[!htb]
    \centering
    \includegraphics[width=.85\linewidth]{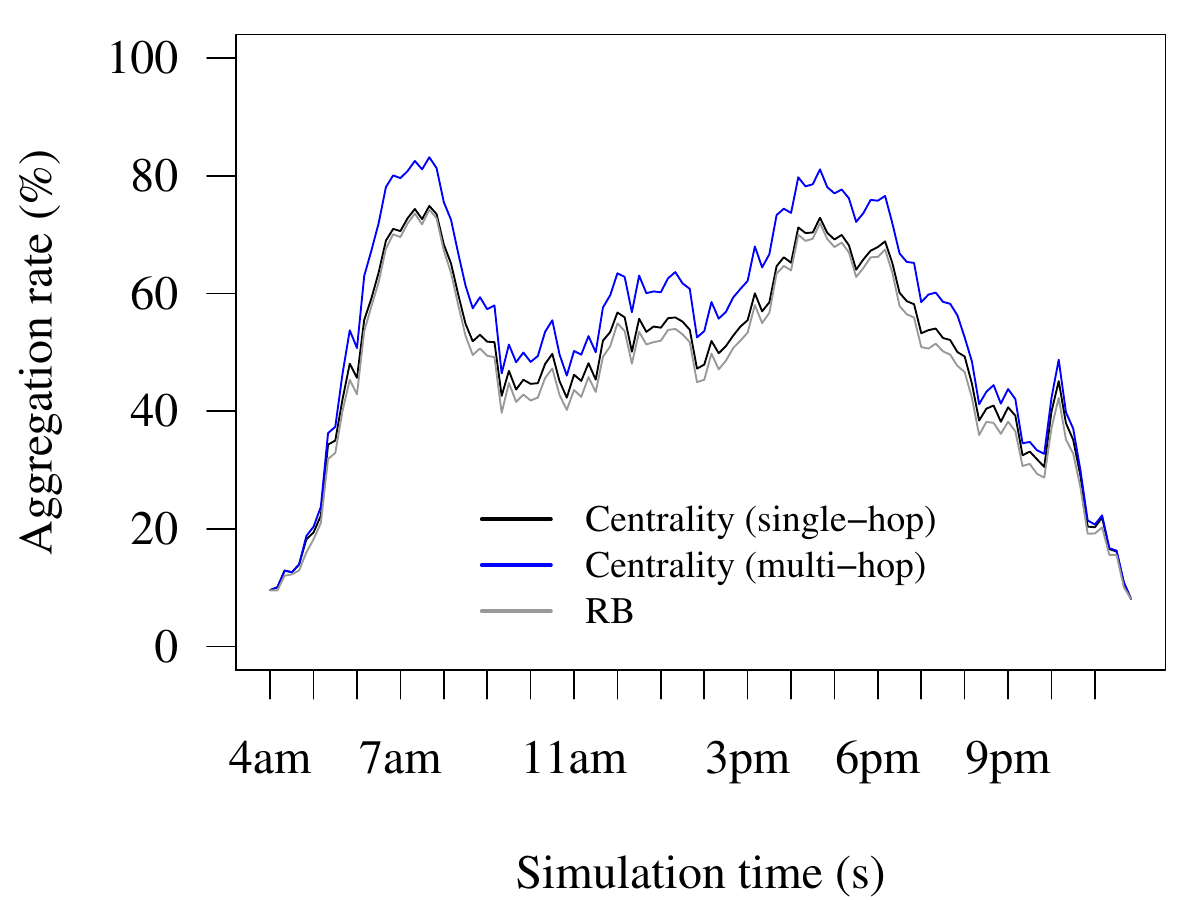}
    \caption{Aggregation rate obtained using the centrality-based algorithm compared to the RB algorithm.}
    \label{fig:aggregation_rate}
\end{figure}

There is an increase in the cost of uploading during peak hours when there is a greater volume of vehicles on the roads. 
In the TAPASCologne data traffic, we observe that an upload cost reached 167.50 kB/s at peak time. 
The high cost of uploading, especially in periods of higher demand, requires offloading techniques to reduce the traffic generated on the network and save bandwidth.
In this way, Figure \ref{fig:aggregation_rate} presents, in detail, the aggregation rate. It corresponds to the ratio between data volume after and before aggregation. The results suggest a significant reduction in sensing data transmitted over the cellular network in both approaches. When we use single-hop communication (1-hop), the Centrality-based algorithm has an aggregation rate close to the RB algorithm. In this scenario, the aggregation point will only collect data from its immediate neighbors.
However, we can improve the aggregation rate by increasing the number of hops. Multi-hop communication (3-hops) enables data collection from more distant vehicles, resulting in a better aggregation rate. The centrality-based algorithm has an aggregation rate equal to 83.14\% in peak hours.

This use case presents a scenario in which the architecture enables the implementation of a solution to select aggregation points and offload sensing data.

\subsection{Mobile Network Application}

The ecosystem of mobile networks is highly dynamic, dense, and connected. As users move between communication technologies, they undergo handover processes, transitioning from one access point or base station (represented by the evolved Node B or eNB) to another \cite{peltonen2021handover}. It is essential to maintain users' connections without any perception of interruption or reduction in transfer rates while they are on the move. Mobile network solutions must ensure that user services are available even in scenarios of high mobility, such as airplanes traveling at speeds of up to 500 km/h \cite{haijun:2017}.

Accurate strategies are essential for transferring and allocating users' resources. Handovers should happen seamlessly and quickly in areas that do not negatively affect mobility and connectivity. Optimizing the allocation of users in mobile network base stations can reduce the handover process. In typical 5G environments, the involvement of multiple stakeholders can hinder the open exchange of information regarding resource availability, operational status, performance capabilities, and service contracts \cite{coronado2022zero}. 
The lack of data sharing is a big challenge that limits the capacity of systems to manage complex chains of end-to-end services. Considering the 5G architectures, these difficulties can be surpassed or minimized through a distributed ledger and an operational data lake \cite{coronado2022zero}. The operational data lake is viewed as a logically centralized repository, as shown in Figure \ref{fig:mobile_app_framework}.

In this scenario, the \texttt{IoT Device} can represent anything connected to the mobile network, such as a vehicle or smartphone. Through the \texttt{DL Interface}, it shares location updates, network service status, or consumes best route directions, optimized connection spots, and general information regarding anomalous events from the Data Lake. The \texttt{Storage System} stores this information. The \texttt{Processing Engine} processes this information and makes it available to mobile users, such as the vehicle.
Additionally, the eNB represents the \texttt{Edge Data} lake.
In this case, the \texttt{Storage System} keeps all data required to network management and optimization. 
The \texttt{Metadata System} enhances and provides additional information to the data in the \texttt{Storage System}. Data stored in data lakes receive little to no processing at the ingestion stage. The goal is to minimize and avoid information loss. Suppose a mobile data service provider realizes it needs to offer a new service or optimize existing ones. The data will be available to the \texttt{Processing Engine}, responsible for data filtering, transformation, aggregation, real-time analytics, event detection, and alerting.
Furthermore, by harvesting the tools provided by the insights layer of the data lake, decision-makers can analyze and compare different models to select solutions while following data governance rules to attend to users' demands. 

\begin{figure}[!htb]
    \centering
    \includegraphics[width=.8\linewidth]{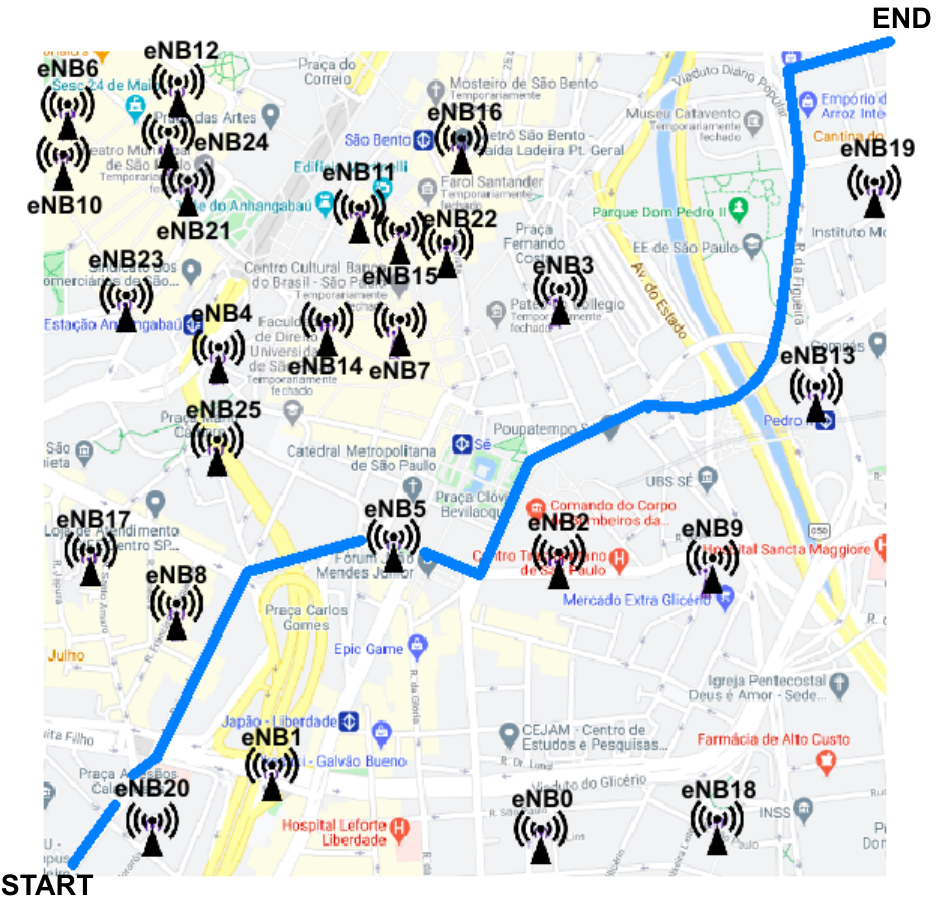}
    \caption{Simulated UE's route at a region in São Paulo city, Brazil, considering real-world eNB locations.}
    \label{fig:spRouteSimulation}
\end{figure}

Once the application and data lake are on, we proceed to show the execution of it in a simulation way.
The objective of our solution is to find the best eNB activation considering the vehicle trajectory.
Figure \ref{fig:spRouteSimulation} presents the mobility scenario of a user that starts at the bottom left and proceeds to the final destination at the top right. The eNB allocation considers the region of São Paulo City, Brazil. The real-world eNB locations are provided by Telebrasil's dataset \cite{telebrasil:2023}. 

The simulation region covers an area of $(1947.65 \times 1878.95)$m$^2$ and comprises 26 eNBs provided by a local phone carrier. Each base station is labeled from 0 to 25. We utilize the SUMO (Simulation of Urban MObility) simulator \cite{Krajzewicz2012SUMO} to generate the route (highlighted in blue in Figure \ref{fig:spRouteSimulation}) that yields 432 UE (User Equipment) location readings. These readings can serve as input for allocation models, such as the ones discussed by Ahmadi et al. \cite{ahmadi2020} and Ramos et al. \cite{ramos20195Gsdn}.

\begin{figure}[!htb]
    \centering
    \includegraphics[width=.7\linewidth]{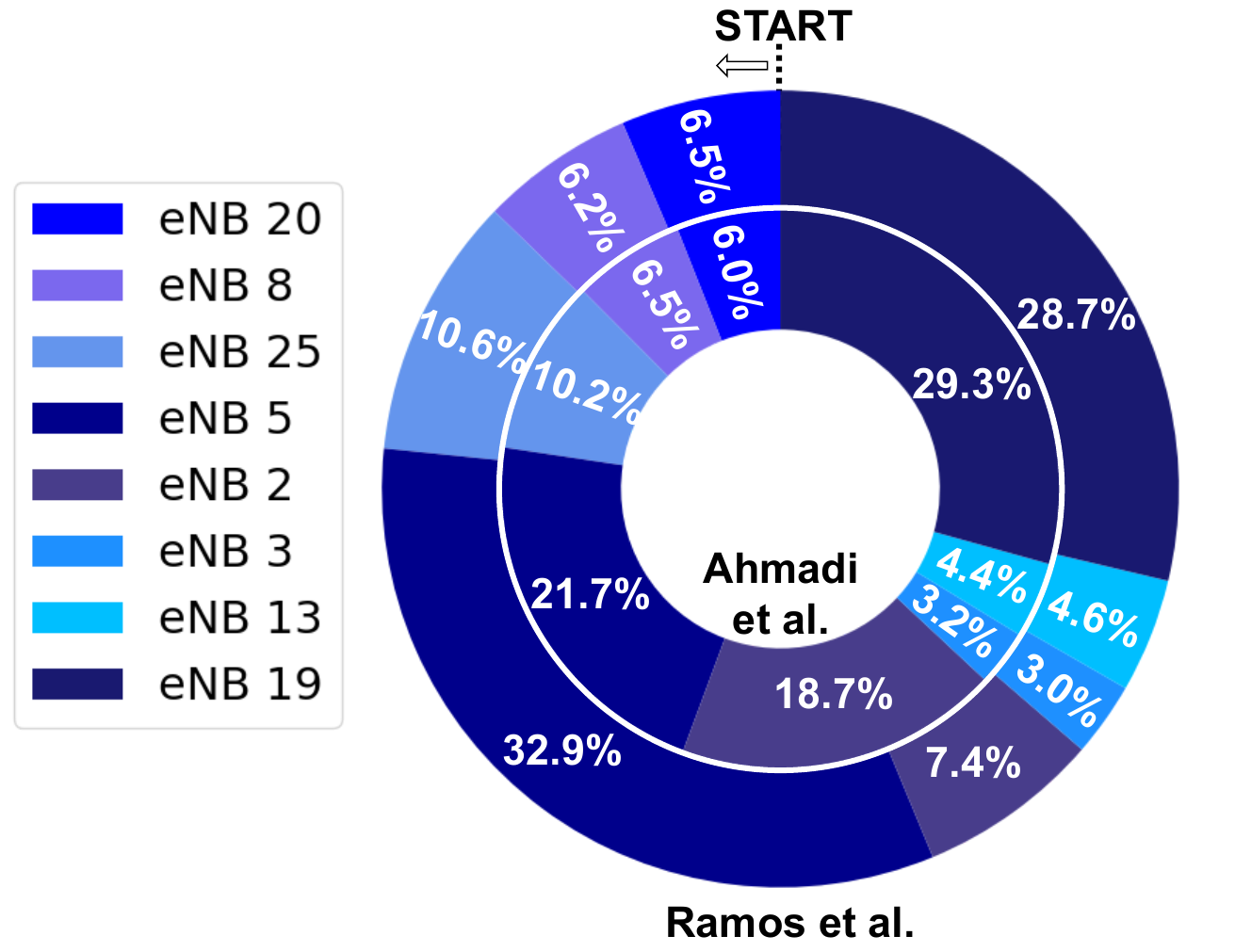}
    \caption{The allocation results for Ramos et al. (outer) and Ahmadi et al. (inner) models.}
    \label{fig:alloResults}
\end{figure}

Figure \ref{fig:alloResults} shows the allocation results for both models (inner results: Ahmadi et al.; outer results: Ramos et al.). We can see that the UE is allocated through the same eNBs and performs 7 handovers, following the allocation sequence 20 -> 8 -> 25 -> 5 -> 2 -> 3 -> 13 -> 19. The models present differences regarding the allocation period along the route. For instance, the model proposed by \cite{ramos20195Gsdn} kept the UE connected to eNB 5 for 32.9\% of the entire route, while \cite{ahmadi2020} kept it connected for 21.9\%. A decision maker can interpret the results and choose models that can keep the UE connected for longer periods in each eNB. Since human mobility is highly predictable \cite{song2010limits}, we can use the data from this route stored in the \texttt{Storage System} of the data lake for the next time the UE starts this specific route. By planning ahead, the data lake can run prediction models through the \texttt{Processing Engine} to reduce the number of handovers by selecting the eNBs with minimal service loss for the user. For example, instead of the 7 handovers initially performed, eNB 25 and eNB 3 can be removed, resulting in the eNB handover sequence 20 -> 8 -> 5 -> 2 -> 13 -> 19.

\subsection{Driver Identification Application}

The problem of driver identification involves classifying drivers based on their behavior, in our case, using machine learning models. We consider a multivariate time series classification problem due to the nature of the data collected from various sensors over time.
%

In this study case, we focus on the feature generation process of machine learning models. 
This process is responsible for generating new features from existing ones to enhance the performance of machine learning models. It involves various techniques, such as mathematical transformations, aggregations, interactions, or domain-specific knowledge.
We propose to use information theory quantifies~\cite{complexity-entropy} as new features in machine learning models. 
Information Theory allows quantifying the information in a dynamic system, such as driver identification. 
Integrating principles from Information Theory into feature generation can achieve a more comprehensive and meaningful representation of the data. These new features enhance machine learning models' capability to handle complexity and extract pertinent knowledge from the data.

We use as features the Complexity Statistical and Entropy measures,
considering a time window of 30 seconds (i.e., a sequence of 30 samples). 
We assumed that each time window was a stationary series. 
For our study, we used a real-world dataset with driving data from four drivers \cite{qar8-sd42-20:2020}. The drivers made several trips using the same vehicle along a fixed path. 
Each record contains 51 features associated with an automotive sensor. 
However, we used only nine features, which are commonly adopted in the literature \cite{uvarov:2021}. 
These features include accelerator pedal value, intake air pressure, absolute throttle position, long-term fuel, engine speed, torque of friction, engine coolant temperature, engine torque, and vehicle speed.

Figure \ref{fig:drive-identification-framework} shows the driver identification framework. 
First, the \texttt{IoT Device} collects the data, represented by smartphones or OBD-II interface.
After that, it transmits to the \texttt{Edge Data Lake} responsible for executing the driver identification models. 
With all features evaluated, we must generate these features and the model to allow this processing. 
To perform this generation, the \texttt{Edge Data Lake} sends the data to the \texttt{Cloud Data Lake}. 
Then, the data is processed, the features extracted, and the machine learning models trained. Finally, the trained model is transferred to the \texttt{Edge Data Lake}, enabling the user to perform low-latency inference tasks.

We compare our proposed features generation, based on Complexity Statistics and Entropy, with the commonly used approaches in the literature \cite{kyung:2019:GAN}. 
We trained seven classifiers using the One-vs-Rest method: 
K-Nearest Neighbors (KNN), 
Linear Support Vector Machine (SVM), 
Radial Basis Function (RBF) SVM, 
Decision Tree, 
Random Forest, 
Multi-Layer Perceptron (MLP), and 
Naive Bayes.
We split the dataset into a training dataset (75\%) and a testing dataset (25\%), using a small dataset and a large dataset with 300 and 9,700 samples from each driver, respectively.

\begin{figure*}[!htb]
\centering
\subfloat[a][Accuracy for small data]{ \includegraphics[width = 0.48\linewidth]{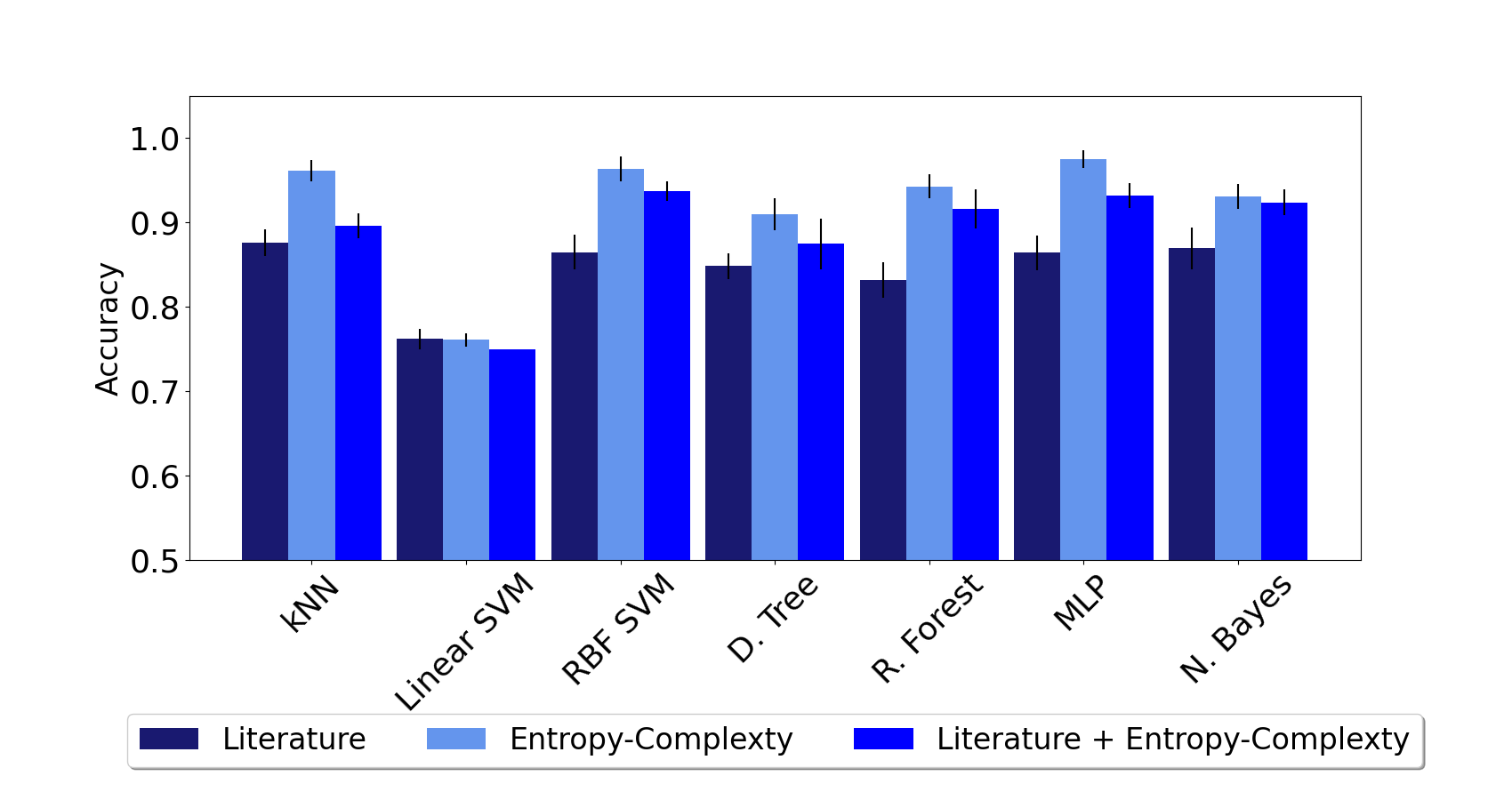} \label{fig:accuracy-small} }%
\subfloat[b][Accuracy for large data]{ \includegraphics[width = 0.48\linewidth]{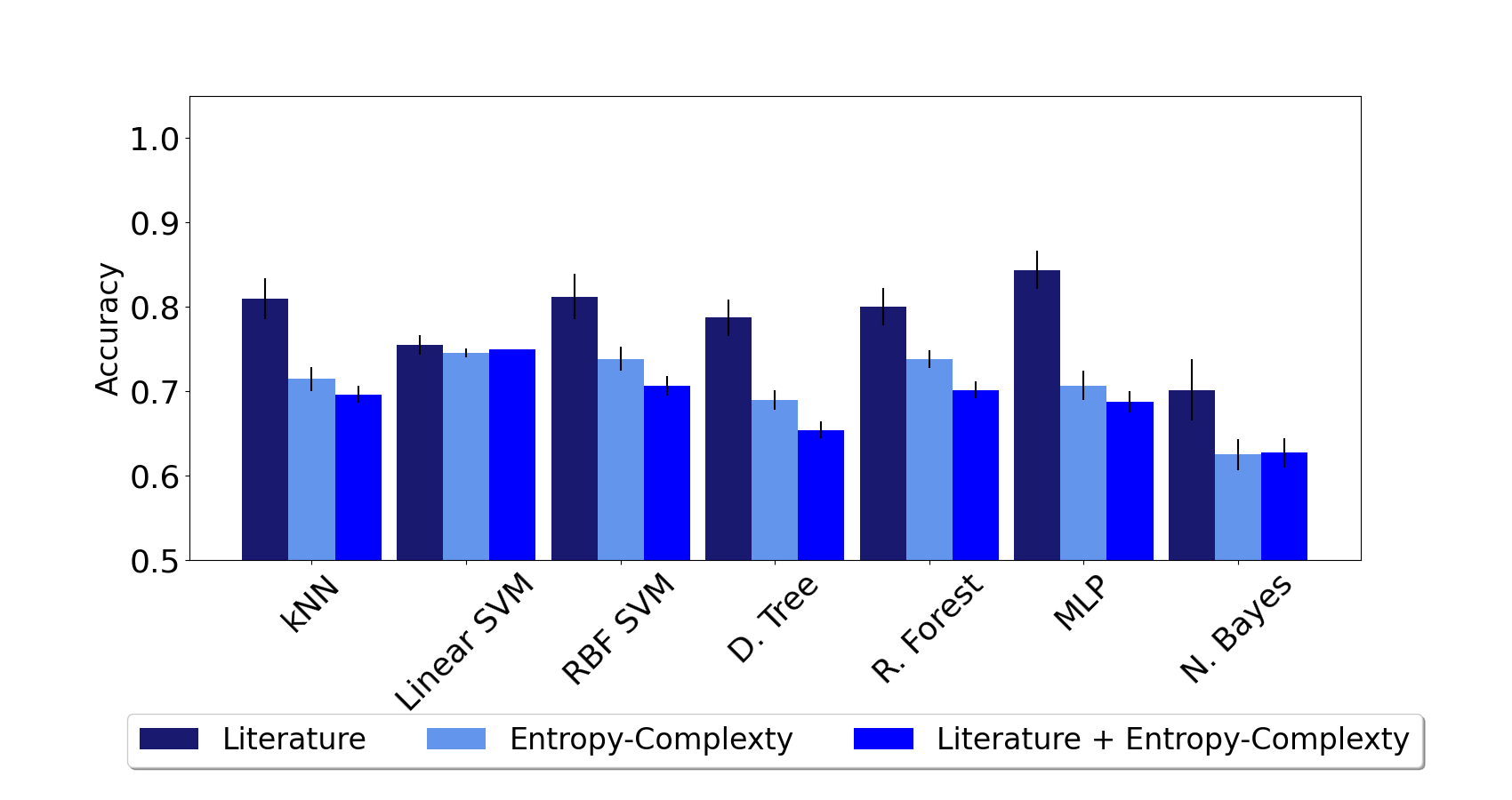} \label{fig:accuracy-large} }
\caption{Classification accuracy results for driver identification.}
\label{fig:driver-identification-results}%
\end{figure*}

Figure \ref{fig:driver-identification-results} presents the accuracy of each classification model. 
The results indicate that the classifier's performance influences the features used and the size of the dataset. 
In the small dataset, we can see that Entropy-Complexity has a lower performance than the other approaches. 
Furthermore, we can significantly improve for most of the used classifiers when we combine them with the pre-processing scheme from the literature. However, in the large data set, the literature approach was equal or better in the used classifiers.
It is noted that all approaches decreased the accuracy for the large dataset compared to the small dataset.

Information theory measures can provide additional features that help to distinguish patterns, behaviors, and classes in time series. Although the results for a large dataset show lower precision than the literature solution in some classifiers, for a small dataset we can obtain a significant improvement. The feature extraction step represents a significant computational overhead in training ML models. However, with the high computing power of \texttt{Cloud Data Lake} or a decentralized solution on \texttt{Edge Data Lake}, we can reduce the workload and improve the efficiency of real-time data processing.

\section{Conclusion} \label{sec:conclusion}

In this work, we presented a distributed Edge-Based Data Lake Architecture for storing and processing ITS data. This architecture is a promising approach to handling different types of data, ranging from automotive to meteorological data, in which the data lake can provide a distributed platform for data integration. The convergence of edge computing and data lake technologies offers significant advantages, such as reducing latency, improving data security, and optimizing bandwidth usage. 

We analyzed three use cases of the proposed architecture.
Besides the data organization and easy access, through well-established interfaces, our qualitative analysis revealed significant improvement in the organization and structure of ITS applications, such as identifying explicitly where the data is collected, stored, and processed. In particular, for the driver identification one, we identify the possibility of increasing the performance using a distributed processing in the \texttt{Edge Data Lake}.
Overall, the proposed architecture represents a valuable contribution to the field of ITS and can pave the way for new advancements in ITS. 
However, some challenges remain to be addressed, such as a data selection algorithm for offloading and a detailed metadata system architecture for complete data integration. 

\begin{acks}
This study was partly financed by the Research Foundation of the State of Alagoas (FAPEAL) under grants E:60030.0000000352/2021 and the National Council for Scientific and Technological Development (CNPq) under grant 407515/2022-4.
\end{acks}

\bibliographystyle{ACM-Reference-Format}
\balance
\bibliography{main}

\end{document}